# Incommensurate Transverse Peierls Transition

F. Z. Yang[1,*], K. F. Luo[2,*], Weizhe Zhang[3], Xiaoyu Guo[3], W. R. Meier[4], H. Ni[5], H. X. Li[1,6], P. Mercado Lozano[7], G. Fabbris[7], A. H. Said[7], C. Nelson[8], T. T. Zhang[9], A. F. May[1], M. A. McGuire[1], R. Juneja[1], L. Lindsay[1], H. N. Lee[1], J.-M. Zuo[5], M. F. Chi[10], X. Dai[11,12], Liuyan Zhao[3], H. Miao[1,†]

[1]*Materials Science and Technology Division, Oak Ridge National Laboratory, Oak Ridge, Tennessee 37831, USA*

[2]*Department of Physics, The University of Texas at Austin, Austin 78712, USA*

[3]*Department of Physics, University of Michigan, Ann Arbor, Michigan 48109, USA*

[4]*Department of Physics and Astronomy, The University of Tennessee, Knoxville, Tennessee 37996, USA*

[5]*Department of Materials Science and Engineering, University of Illinois at Urbana-Champaign, Urbana, Illinois 61801, USA*

[6]*Advanced Materials Thrust, The Hong Kong University of Science and Technology (Guangzhou), Guangzhou, Guangdong 511453, China.*

[7]*Advanced Photon Source, Argonne National Laboratory, Argonne, Illinois 60439, USA*

[8]*National Synchrotron Light Source II, Brookhaven National Laboratory, Upton, New York 11973, USA*

[9]*Institute of Theoretical Physics, Chinese Academy of Sciences, Beijing 100190, China*

[10]*Center for Nanophase Materials Sciences, Oak Ridge National Laboratory, Oak Ridge, Tennessee 37831, USA*

[11]*Materials Department, University of California, Santa Barbara, California 93106, USA*

[12]*Department of Physics, Hong Kong University of Science and Technology, Clear Water Bay, Hong Kong*

**In one-dimensional quantum materials, conducting electrons and the underlying lattices can undergo a spontaneous translational symmetry breaking, known as Peierls transition. For nearly a century, the Peierls transition has been understood within the paradigm of electron-electron interactions mediated by longitudinal acoustic phonons. This classical picture has recently been revised in topological semimetals, where transverse acoustic phonons can couple with conducting *p*-orbital electrons and give rise to an unconventional Fermi surface instability, dubbed the transverse Peierls transition (TPT). Most interestingly, the TPT induced lattice distortions can further break rotation or mirror/inversion symmetries, leading to nematic or chiral charge density waves (CDWs). Quantum materials that host the TPT, however, have not been experimentally established. Here, we report the experimental discovery of an incommensurate TPT in the tetragonal Dirac semimetal EuAl$_4$. Using**

**inelastic x-ray scattering with meV resolution, we observe the complete softening of a transverse acoustic phonon at the CDW wavevector upon cooling, whereas the longitudinal acoustic phonon is nearly unchanged. Combining with first principles calculations, we show that the incommensurate CDW wavevector matches the calculated charge susceptibility peak and connects the nested Dirac bands with Al $3p_x$ and $3p_y$ orbitals. Supplemented by second harmonic generation measurements, we show that the CDW induced lattice distortions break all vertical and diagonal mirrors whereas the four-fold rotational symmetry is retained below the CDW transition. Our observations strongly suggest a chiral CDW in EuAl$_4$ and highlight the TPT as a new avenue for chiral quantum states.**

The charge density wave (CDW), a spontaneous translational symmetry breaking in electron liquids, plays a central role in many correlated and topological materials, including high-$T_c$ superconductors[1-3], kagome metals[4-8], axion insulators[9,10], and quantum Hall liquids[11-15]. Historically, the first microscopic mechanism of CDW was described by the Fröhlich model[16], where electron-electron scatterings mediated by longitudinal acoustic (LA) phonons are divergent at the Fermi surface nesting vector, $Q_{CDW}$. Consequently, the formation of a CDW accompanied by softening of LA phonons, known as the Kohn anomaly[17], and longitudinal lattice distortions as schematically shown in Figs. 1**a** and 1**b**. Recently, it was predicted that topological electrons with anisotropic $p$-orbital characters can strongly couple to transverse acoustic (TA) phonons, leading to CDW instabilities and TA Kohn anomaly, named the transverse Peierls transition (TPT)[13]. This new mechanism is shown schematically in Fig. 1**c** and 1**d**. Interestingly, since transverse lattice distortions are two-component vectors, TPTs can lead to unconventional CDWs such as nematic (Fig. 1**e**) and chiral (Fig. 1**f**) CDWs. The TPT was originally proposed to explain the 3-dimensional quantum Hall effect in the Dirac-semimetal ZrTe$_5$[13-15]. However, the TPT induced CDW superlattice peaks in ZrTe$_5$ have not been observed. Recently, x-ray scattering studies[18-22] suggest the possible realization of TPT in a topological magnet EuAl$_4$ that is reported to feature the spontaneous chirality flipping[20]. In this letter, we establish the incommensurate TPT and show compelling experimental signatures of a novel chiral CDW in EuAl$_4$.

EuAl$_4$ has a tetragonal structure with space and point groups of *I4/mmm* and *4/mmm* (D$_{4h}$), respectively. At room temperature, EuAl$_4$ is a semi-Dirac topological semimetal[20,23,24], which has a linear dispersion along the $k_z$ direction and a quadratic dispersion along the $k_x/k_y$ direction.

Below the critical temperature $T_{CDW}$=141 K, EuAl$_4$ forms an incommensurate CDW with $\mathbf{Q}_{CDW}$=(0, 0, 0.183) in reciprocal lattice units (r.l.u.) along the $k_z$ direction[18-22]. The $\mathbf{Q}_{CDW}$ connects the semi-Dirac bands[20], suggesting a Fermi surface nesting assisted CDW. Figure 1**g** shows the temperature dependence of the CDW superlattice peak at $\mathbf{Q}$=(2, 2, 2)-$\mathbf{Q}_{CDW}$. Below $T$=15.4 K, EuAl$_4$ further experiences multiple spin density wave (SDW) transitions including the emergence of chiral SDW and spontaneous chirality flipping[20,25,26].

We first examine the temperature dependent phonon dynamical structure factor, $S(\mathbf{Q}, \omega)$, of the LA and TA modes in the vicinity of $\mathbf{Q}_{CDW}$. Figure 2**a-d** show $S(\mathbf{Q}, \omega)$ at two selected temperatures, 300K and 160K, and along two momentum cuts, (2, 2, L∈ [1.5, 2.0]) and (0, 0, L∈ [7.5, 8.0]), measured by the inelastic x-ray scattering (IXS). The cross-section of IXS is proportional to $(\sum_j \mathbf{Q} \cdot \mathbf{u}_j)$, where $\mathbf{Q}=\mathbf{k}_f-\mathbf{k}_i$ is the momentum transfer of incident ($\mathbf{k}_i$) and scattered ($\mathbf{k}_f$) photons, and $\mathbf{u}_j$ is the distortion of atom-$j$ in the unit cell (See Supplementary Materials Section I). The $S(\mathbf{Q}, \omega)$ spectra along the (2, 2, L∈ [1.5, 2.0]) and (0, 0, L∈ [7.5, 8.0]) directions thus select the TA and LA modes, respectively. By comparing magnified $S(\mathbf{Q}, \omega)$ maps near $\mathbf{Q}_{CDW}$ shown in Figs. 2**a** and 2**b**, we observe softening of TA phonons from 300 K to 160 K. This observation is in stark contrast to the LA mode that slightly hardens, as expected in a harmonic approximation on cooling (Figs. 2**c** and 2**d**). Figure 2**e** summarizes the temperature dependent evolution of extracted TA and LA phonon dispersions. Representative IXS spectra at $\mathbf{Q}$=(2, 2, 1.83) are shown in Fig. 2**f**, highlighting the softening of TA phonons. Figure 2**g** shows the extracted TA phonon softening, $\Delta E (T)=\omega_0(\mathbf{Q})-\omega(\mathbf{Q},T)$, at $T$ = 300, 160, and 130 K. Here $\omega_0(\mathbf{Q})$ denotes the sinusoidal bare TA phonon band. The sharp peak of $\Delta E (T)$ at $\mathbf{Q}_{CDW}$ establishes the TA Kohn anomaly and hence strongly supports a TPT.

To uncover the origin of the TA Kohn anomaly, we calculate the electronic band structure and momentum dependent charge susceptibility. Although the Eu $f$-electrons are far away from the Fermi level[20], the large magnetic moment and complex magnetic ground state of EuAl$_4$ have non-trivial effect on the DFT calculated lattice dynamics (See Supplementary Materials Section III). We therefore performed calculations on SrAl$_4$ with nearly identical electronic structure near the Fermi level (see Supplementary Materials Section III) and a CDW instability (see Supplementary Materials Section I). As shown in Fig. 3**a**, the heavily hybridized Al $p$-orbitals form a semi-Dirac point above the $E_F$. A nesting of the semi-Dirac bands produce a charge susceptibility peak at $\mathbf{Q}$~(0, 0, 0.12) for SrAl$_4$ as is shown in Fig. 3**b**. Encouragingly, the calculated phonon spectrum displays a negative energy for the TA and a strong electron-TA phonon coupling. The momentum $q$, at which the TA mode reaches its minimum shown in Figs. 3c and 3d, matches well the calculated charge susceptibility and is consistent with the experimentally determined $Q_{CDW}$~0.125 in SrAl$_4$ (see Supplementary Materials Section I). These first-principles calculations together with the experimental observation of the TA Kohn anomaly (Fig. 2) provide compelling evidence of a TPT induced CDW.

Having the incommensurate TPT established, we then proceed to reveal the possible chiral nature of the CDW by investigating the broken point symmetries using a combination of elastic x-ray scattering and high-sensitivity rotation anisotropy second harmonic generation (RA-SHG). As schematically illustrated in Figs. 4a and 4b, the achiral nematic CDW breaks the $C_4$ rotational symmetry, whereas the chiral CDW breaks the mirror/inversion symmetries but retains the $C_4$ rotational symmetry (see Supplementary Materials Section I). We start with the elastic x-ray scattering data shown in Figs. 4c and 4d which closely track the temperature dependence of two 90°-rotated Bragg peaks, (1, 0, 7) and (0, 1, 7), respectively. Across a wide temperature range including $T_{CDW}$, we observe consistently narrow half-widths-at-half-maximum (HWHM) of both Bragg peaks (HWHM~0.0025 r.l.u) without splitting or broadening. This result puts an upper limit for the rotational symmetry breaking lattice distortions to be on the order of 0.1% of the lattice constant, i.e., < 1pm, smaller than the fundamental atomic position uncertainty ~ 2pm derived from the Heisenberg uncertainly principle (see Methods). Therefore, our x-ray measurements exclude the rotational symmetry breaking for the CDW phase. Indeed, the $C_4$ rotational symmetry breaking is only observed in the helical SDW phase transition below 13 K (See Supplementary Materials Section I), consistent with a $C_4$-preserving CDW.

We then move to the RA-SHG results. Due to the $C_4$ and $C_2$ rotational (screw) symmetry along the c-axis, there should be no SHG signal in the normal incidence geometry, thus we need to perform RA-SHG measurements in the oblique incidence geometry for EuAl$_4$ with the (001) facet (see Methods and Supplementary Materials Section II)[27,28]. Figure 4e shows the RA-SHG polar plots, i.e., the SHG intensity as a function of the rotating angle $\phi$ between the light scattering plane and the crystalline a-axis, with both fundamental and SHG polarizations chosen to be parallel to the scattering plane ($P_{in}$-$P_{out}$), at four selected temperatures: 80K and 135K (below $T_{CDW}$), and 155K and 220K (above $T_{CDW}$). We can see the presence of $C_4$ in RA-SHG for all temperatures and the non-monotonic change in the SHG intensity upon cooling across $T_{CDW}$. We have identified electric quadrupole as the main SHG radiation source and have ruled out other possibilities (see Supplementary Materials Section II).

To quantify the point symmetry evolution across $T_{CDW}$, we fit our RA-SHG polar plot at every temperature to extract the isotropic (*i.e.*, a constant), the $C_2$ symmetric (*i.e.*, $\propto \cos(2\phi)$), and the $C_4$ symmetric (*i.e.*, $\propto \cos(4\phi)$) components, as well as its orientation away from the a-axis (*i.e.*, $\phi_0$) (see Supplementary Materials Section II), and show their temperature dependencies in Figs. **4f** and **4g**. First, we note that the isotropic and the $C_4$ components are much stronger than those for $C_2$, and furthermore, that the strong isotropic and $C_4$ signals exhibit a similarly dramatic temperature dependence with their intensities peaking around $T_{CDW}$ (Fig. **4f**). The weak $C_2$ signal, on the other hand, is nearly temperature independent. We attribute this $C_2$ signal as an experimental systematic background based on its small amplitude and temperature independence across the phase transition. The absence of $C_4$ symmetry breaking is consistent with the elastic x-ray scattering results in Figs. **4c** and **4d**. Second, we notice that the RA-SHG rotates away from alignment with the crystal axis as the temperature decreases below $T_{CDW}$ (Fig. **4g**). This orientation change, although small, is direct evidence of breaking the in-plane $C_2$ rotational symmetries, the vertical mirror symmetries, and the diagonal mirror symmetries for the CDW phase, as it means a absence of any special in-plane direction to which the RA-SHG pattern locks to below $T_{CDW}$ (see Supplementary Materials Section II for the same results from RA-SHG in other polarization channels)[29-32]. Furthermore, this rotation of the RA-SHG pattern shows that the CDW state has handedness, being ferro-rotational or chiral.

Although TPT can host both nematic and chiral CDWs as shown in Fig. 1, the $C_4$ symmetric CDW, as identified in EuAl$_4$, can be energetically favored by considering the form factor of the CDW gap in the electronic structure (See Supplementary Materials III). As shown in Figs. **4a** and **4b**, the $C_2$ nematic CDW gives rise to an anisotropic CDW gap and a TA Kohn anomaly along either (1, 0, 0) or (1, 1, 0) directions, whereas the $C_4$ chiral CDW results in an isotropic CDW gap and the TA Kohn anomalies along both (1, 0, 0) and (1, 1, 0) directions. Indeed, the soft TA mode shown in Fig. 2 corresponds to a transverse lattice instability along the (1, 1, 0) direction. At the same time, the transverse lattice instability along the (1, 0, 0) direction has also been observed in EuAl$_4$,[21] in agreement with the $C_4$ lattice distortions with broken mirrors. We note that, while our macroscopic x-ray and optical probes cannot distinguish ferro-rotational CDW (point group 4/m) and chiral-CDW (point group 4) with structural twinning, the incommensurability of the CDW naturally favors the chiral CDW.

In summary, we observed incommensurate TPT in the Dirac semimetal EuAl$_4$. The emergence of handed lattice distortions highlights the rich landscape of TPT in electron-TA phonon coupled topological materials.

## Methods

**Sample preparation:**

High-quality EuAl4 crystals were grown from an aluminum flux, with their growth and characterization described in Ref. 26.

**X-ray scattering measurements:**

The single crystal elastic X-ray diffraction was performed at the 4-ID-D beamline of the Advanced Photon Source (APS), Argonne National Laboratory (ANL), and the integrated *in situ* and resonant hard X-ray studies (4-ID) beam line of National Synchrotron Light Source II (NSLS-II). The photon energy, which is selected by a cryogenically cooled Si(111) double-crystal monochromator, is 6.977 keV (Eu $L_3$ resonance).

**The 4-ID-D, APS**: the X-rays higher harmonics were suppressed using a Si mirror and by detuning the Si (111) monochromator. Diffraction was measured using a vertical scattering plane geometry and horizontally polarized (σ) X-rays. The incident intensity was monitored by a $N_2$ filled ion chamber, while diffraction was collected using a Si-drift energy dispersive detector with approximately 200 eV energy resolution. The sample temperature was controlled using a He closed cycle cryostat and oriented such that X-rays scattered from the (001) surface.

**The 4-ID, NSLS2**: The sample is mounted in a closed-cycle displex cryostat in a vertical scattering geometry. The incident X-rays were horizontally polarized, and the diffraction was measured using a silicon drift detector.

**Inelastic X-ray scattering measurements:**

The experiments were conducted at beam line 30-ID-C (HERIX) at the APS. The highly monochromatic X-ray beam of incident energy $E_i$ = 23.7 keV was focused on the sample with a beam cross section of ~35 × 15 mm$^2$ (horizontal × vertical). The total energy resolution of the monochromatic X-ray beam and analyzer crystals was ΔE~1.3 meV (full width at half maximum). The measurements were performed in transmission geometry. Typical counting times were in the range of 30 – 120 s per point in the energy scans at constant momentum transfer **Q**.

**Second harmonic generation measurements:**

RA-SHG measurements are conducted using an 800 nm wavelength ultrafast laser source, with a repetition rate of 200 kHz and a pulse width of 70 fs. The fundamental light shines onto the sample in an oblique incidence geometry with an incident angle $\theta = 11.4°$. After passing through an achromatic lens (30 mm focal length), the laser beam is focused onto the sample surface with a full width at half maximum (FWHM) of 30 $\mu$m and a fluence of approximately 1.4 mJ/cm$^2$. A half-wave plate and a polarizer are used to select the *P*- or *S*-polarized component of the incident fundamental light ($P/S_{in}$) and the reflected SHG light ($P/S_{out}$), allowing for measurements in four channels: $P_{in}$-$P_{out}$, $P_{in}$-$S_{out}$, $S_{in}$-$P_{out}$ and $S_{in}$-$S_{out}$ as a function of the rotation angle of the scattering plane. Here *P* and *S* represent photon polarization parallel and perpendicular to the scattering plane, respectively. A set of filters have been used to remove the fundamental 800 nm light before the reflected light being collected by a charge-coupled device. The experiments are conducted in an environment with a pressure less than $1 \times 10^{-6}$ mbar with a base temperature of 80 K.

**Scanning transmission electron microscopy measurements:**

TEM specimen was prepared by mechanical polishing and ion milling to electron transparency. STEM experiments were performed on a JEOL JEM-ARM200F (NeoARM) aberration-corrected scanning transmission electron microscope (STEM) equipped with a custom-built Gatan double-tilt liquid nitrogen cooling holder. The base temperature of our cooling holder is ~92.5 K and the holder could be stabilized for nearly eight hours per liquid nitrogen filling. Annular dark field images were acquired with 27.4 mrad convergence semiangle and 68-280 mrad collection angle. At 93 K, image series containing 30 fast-scanning, low signal-to-noise ratio atomic resolution images were registered and averaged to produce high quality atomic resolution STEM image.

**First principles calculations:**

**Electronic structures:** Ab initio calculations were carried out for EuAl$_4$ and SrAl$_4$ (no.139 space group I4/mmm), using the relaxed lattice parameters of the primitive cell. We used DFT within the generalized gradient approximation of Perdew, Burke, and Ernzerhof[33]. The core–valence interaction was described by means of norm-conserving pseudopotentials[34]. Electron wavefunctions were expanded in a plane waves basis set with a kinetic energy cutoff of 68Ry to achieve energy accuracy better than 1meV per atom, and the Brillouin zone was sampled using a 11×11×11 Γ-centered Monkhorst–Pack mesh. All DFT calculations were performed using the Quantum ESPRESSO package[35,36]. We adopt a degauss σ = 1mRy of Fermi-Dirac smearing function.

To inspect nesting strength of different momenta on the Fermi surface, we calculated Fermi surface nesting functions[37] $\chi_0''(\boldsymbol{q}) = \sum_k \delta(\epsilon_{\boldsymbol{k}} - \epsilon_F)\delta(\epsilon_{\boldsymbol{k+q}} - \epsilon_F)$, which are the imaginary part of the static susceptibility functional with the double-delta approximation, and the Dirac-delta function is replaced by a Lorentzian of width 10meV in the numerical implementation. 16 electronic bands near the Fermi level were interpolated by using maximally localized Wannier functions, to a 100×100×100 k-grid, to achieve accurate calculation of Fermi surface nesting. The peak at Q = 0.137 (2π/c) provides evidence to connect the lattice instability with Fermi surface nesting.

**Phonon band structure and electron-phonon coupling:** Lattice-dynamical properties were calculated using density functional perturbation theory (DFPT) and performed using the Quantum ESPRESSO package. The soft TA modes are found to be centered around Q = 0.1(2π/c), in agreement with the nesting effect. Since the imaginary frequencies of these soft modes are small, 0.25THz, and they only appear after interpolation from a 3x3x3-grid to the finer grid, an additional DFPT calculation at the exact Q wave vector was performed to confirm the lattice instability[38].

Calculations of electron–phonon couplings were performed using the EPW code[39]. To evaluate phonon self-energy and linewidth, we computed electronic and vibrational states as well as their coupling g-matrix on a 11×11×11 and 3×3×3 Brillouin-zone grid, respectively. Based on the Wannier functions, the g-matrix is interpolated to the high-symmetry Γ-Z path, which is used to estimate phonon linewidths from the imaginary parts of phonon self-energies: $\Pi''_{qv}(\omega, T) = 2\pi \sum_{mnk} |g_{mnv}(\boldsymbol{k}, \boldsymbol{q})|^2 [f_T(\varepsilon_{nk}) - f_T(\varepsilon_{mk+q})] \delta(\varepsilon_{mk+q} - \varepsilon_{nk} - \omega)$. Note that temperature $T$ was set to be 100 K in the calculation.

**Quantum Uncertainty of the Lattice Distortion:**

The Heisenberg uncertainty principle states that: $\Delta x \cdot \Delta p \geq \frac{h}{4\pi}$, where $h$ is the Planck constant. $\Delta p \leq \sqrt{2 m_{Al} E}$, where $m_{Al}$=27 u is the atomic mass of Aluminum atom, and $E \sim 50$ meV is the highest phonon energy. Therefore, the upper limit of $\Delta x \geq \frac{h}{4\pi \sqrt{2 m_{Al} E}} \sim 2$ pm.

**Acknowledgements**

We thank Juba Bouaziz, Stefan Blügel, Cristian Batista, Andrew Christianson, Satoshi Okamoto, and Yang Zhang for stimulating discussions. This research was supported by the U.S. Department of Energy, Office of Science, Basic Energy Sciences, Materials Sciences and Engineering Division (X-ray, sample growth, and theory). X-ray scattering used resources (beamline 4ID and 30ID) of the Advanced Photon Source, a U.S. DOE Office of Science User Facility operated for the DOE Office of Science by Argonne National Laboratory under Contract No. DE-AC02-06CH11357. X-ray scattering measurements used resources at 4-ID of the National Synchrotron Light Source II, a US Department of Energy Office of Science User Facility operated for the DOE Office of Science by Brookhaven National Laboratory under contract no. DE-SC0012704. Microscopy was supported by a DOE-BES Early Career project FWP #ERKCZ55 (H.N.). Technique development was performed under DOE, Basic Energy Sciences, Materials Sciences, and Engineering Division (M.C.) and microscopy performed at the ORNL's Center for Nanophase Materials Sciences (CNMS), which is a DOE Office of Science User Facility.

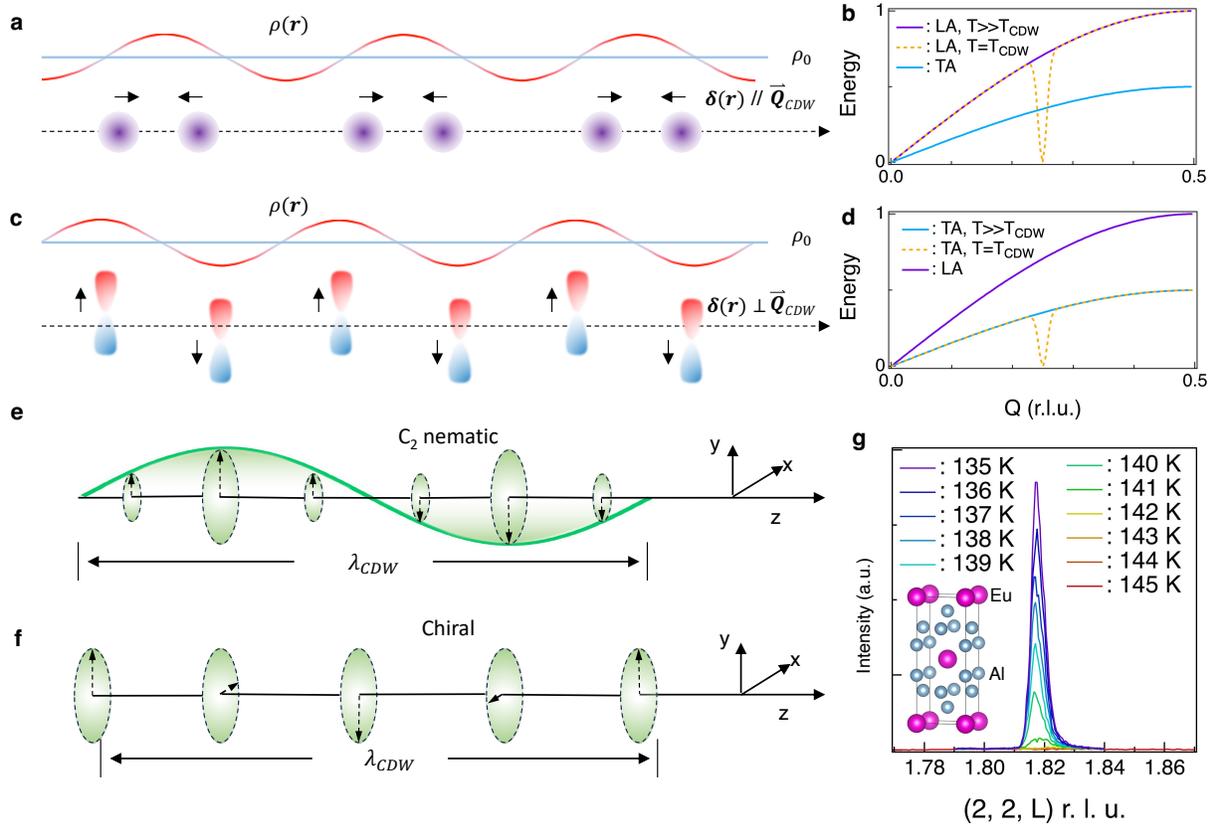

**Figure 1: Longitudinal and transverse Peierls transitions. a**, Schematic of a classical Peierls transition in 1D. The interaction between a LA phonon and an isotropic *s*-orbital conduction electrons yields a CDW instability. **b**, The LA Kohn anomaly induced by the classical Peierls transition. **c**, Transverse Peierls transition driven by the coupling between conduction electrons with anisotropic orbital characters, such as *p*-orbitals, and TA phonons. **d**, The TA Kohn anomaly induced by the TPT. Distinct from the LA wave, the TA wave is a vector wave (similar to an electromagnetic wave) and hence can carry finite angular momentum. **e** and **f** depict linearly polarized and circularly polarized TA waves, respectively. The linearly polarized lattice distortion breaks (screw) rotational symmetry, $C_n$, giving rise to a $C_2$ nematic CDW. The circular polarized lattice distortion preserves (screw) rotational symmetry, $C_n$, and can host a chiral CDW. **g**, Temperature dependence of the incommensurate-CDW superlattice peak at (2, 2, 2-0.183) showing a CDW transition at $T_{CDW}$=142 K. Inset shows the crystal structure of EuAl$_4$. Magenta and silver circles represent Eu and Al atoms, respectively.

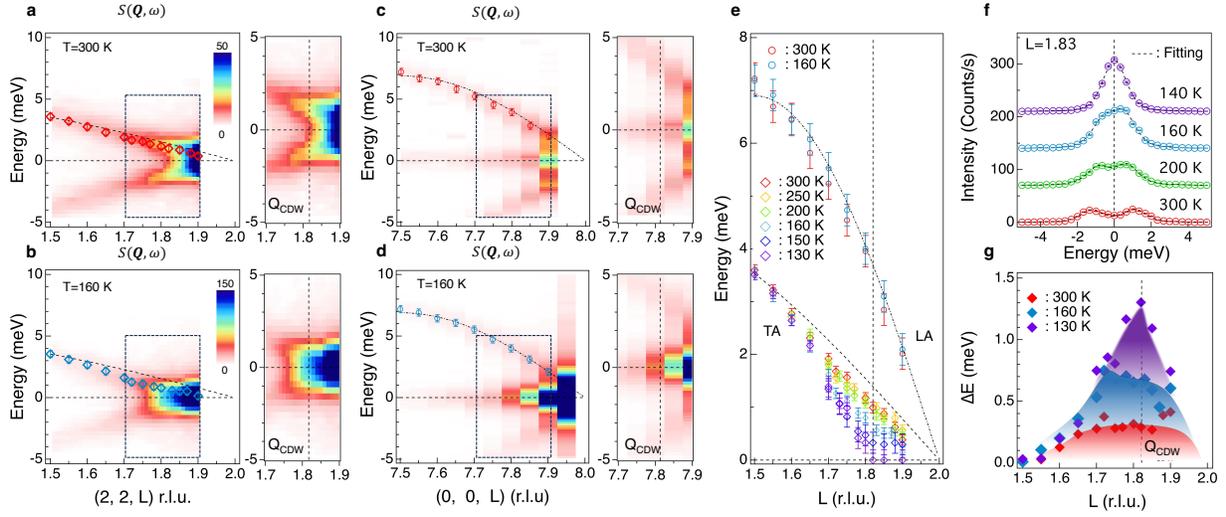

**Figure 2: Observation of TA Kohn anomaly. a** and **b** show TA phonon dynamical structure factor, $S(\mathbf{Q}, \omega)$ along the (2, 2, L∈ [1.5, 2.0]) direction, at T=300 and 160 K, respectively. Red and cyan diamonds are extracted phonon peak positions at 300 and 160 K, respectively. The dashed curves represent the sinusoidal bare band of the TA phonon. Right panels in **a** and **b** show magnified view of $S(\mathbf{Q}, \omega)$ near $Q_{CDW}$. **c** and **d** are the same plots as **a** and **b** but along the (0, 0, L∈ [7.5, 8.0]) direction to select the LA phonon branch. **e**, Temperature dependence of extracted TA and LA phonon dispersions. The TA phonon modes show a large energy renormalization whereas the LA phonon is slightly hardened. **f**, Temperature dependent IXS spectra at Q=(2,2,1.83). **g**, TA phonon softening, $\Delta E(T)$ at 300K, 160K and 130K, relative to the bare band dispersion (dashed sinusoidal curves **a**).

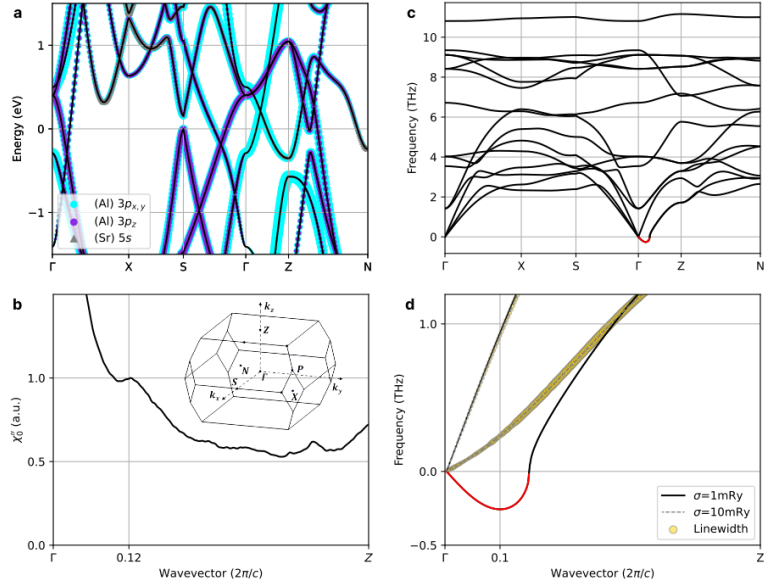

**Figure 3: DFT calculation of electron-TA phonon coupling driven TPT. a,** Electronic band structure of non-magnetic sister compound SrAl$_4$. Light blue, purple, and grey marks represent Al $3p_{x,y}$-orbitals, Al $3p_z$-orbital, and Sr $5s$-orbital, respectively. The Brillouin zone and high symmetry point in momentum space are shown in **b**. The "nesting" of the linearly dispersed semi-Dirac bands give rise to a peak in the calculated charge susceptibility, $\chi_0''$. The susceptibility peak position matches the calculated lattice instability of the TA phonon shown in **c** and **d**. Dashed grey curve and solid black curves are calculated for different electron temperatures. The negative TA phonon at σ=1 mRy supports electron-TA phonon coupling driven TPT and is consistent with experimental observations.

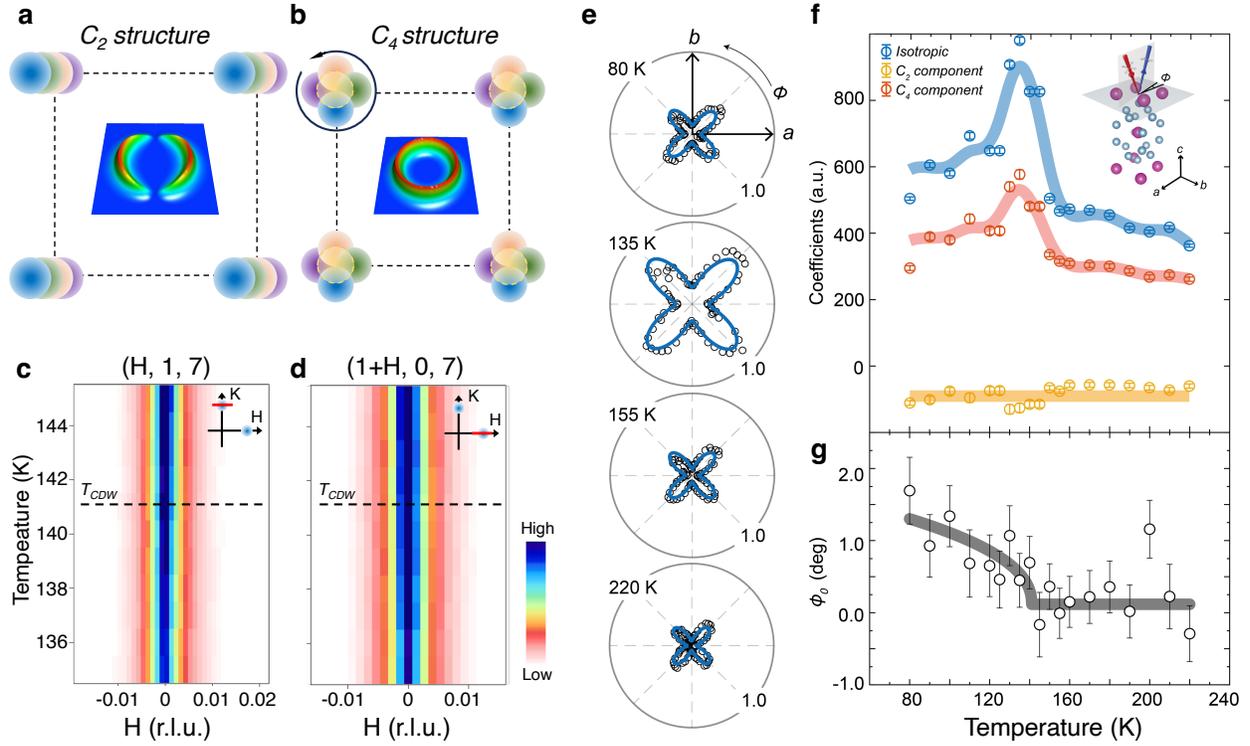

**Figure 4. Signature of TPT induced chiral CDW in EuAl$_4$**. **a** and **b**, Two possible lattice distortion patterns that host C$_2$ nematic and C$_4$ chiral CDWs. Since the TA softening selects a specific direction in the C$_2$ nematic CDW, the CDW gap will take a *p*-wave like gap function where the gap minimums are in the direction that is perpendicular to the nematic direction. The screw C$_4$-preserving chiral CDW induces TA softening along all high symmetry directions and hence yields a *s*-wave gap function. **c** and **d**, temperature dependence of the C$_4$-related fundamental Bragg peaks (0, 1, 7) and (1, 0, 7) across $T_{CDW}$. **e**, RA-SHG patterns at 80 K, 135 K, 155 K, and 220 K. Open circles are measured data, the blue solid lines are the best fits. Numbers at the corners are scales of the plots, with 1.0 representing 6 counts of the SHG photons per second per microwatt. **f**, Temperature-dependence of fitted coefficients from point group *mmm*. Open circles are measured data with error bars being one standard deviation. Solid lines are guides to the eyes. Inset: SHG in the oblique incidence geometry overlaid on the crystal structure of EuAl$_4$. **g**, Temperature-dependence of rotation angle $\phi_0$ from point group *4/m* or *4*. Open circles are measured data with error bars being one standard deviation and the grey curve is a guide to the eyes.